# Strong magnetoelectric coupling in mixed ferrimagnetic-multiferroic phases of a double perovskite

M. K. Kim, J. Y. Moon, S. H. Oh, D. G. Oh, Y. J. Choi[*], and N. Lee[*]

*Department of Physics, Yonsei University, Seoul 03722, Korea*

Exploring new magnetic materials is essential for finding advantageous functional properties such as magnetoresistance, magnetocaloric effect, spintronic functionality, and multiferroicity. Versatile classes of double perovskite compounds have been recently investigated because of intriguing physical properties arising from the proper combination of several magnetic ions. In this study, it is observed that the dominant ferrimagnetic phase is coexisted with a minor multiferroic phase in single-crystalline double-perovskite $Er_2CoMnO_6$. The majority portion of the ferrimagnetic order is activated by the long-range order of $Er^{3+}$ moments below $T_{Er}$ = 10 K in addition to the ferromagnetic order of $Co^{2+}$ and $Mn^{4+}$ moments arising at $T_C$ = 67 K, characterized by compensated magnetization at $T_{Comp}$ = 3.15 K. The inverted magnetic hysteresis loop observed below $T_{Comp}$ can be described by an extended Stoner–Wohlfarth model. The additional multiferroic phase is identified by the ferroelectric polarization of ~0.9 $\mu C/m^2$ at 2 K. The coexisting ferrimagnetic and multiferroic phases appear to be strongly correlated in that metamagnetic and ferroelectric transitions occur simultaneously. The results based on intricate magnetic correlations and phases in $Er_2CoMnO_6$ enrich fundamental and applied research on magnetic materials through the scope of distinct magnetic characteristics in double perovskites.

Correspondence and requests for materials should be addressed to Y. J. C. (phylove@yonsei.ac.kr) or N. L. (eland@yonsei.ac.kr).

**Introduction**

One of the ideas behind examining magnetic materials aims to develop desired functional properties utilized in a wide range of technologies, for example, energy storage, [1] memory devices [2], medical appliances [3], and environmental monitoring sensors [4]. In particular, magnetic oxides comprising metal cations and oxygen anions have been extensively investigated owing to the abundance of the constituents and stability of the compounds. A prominent example can be found in perovskite rare-earth manganites that have been the focus of research on magnetic materials over the last few decades. In mixed-valence manganites, the subtle balance between hopping and localization of charge carriers leads to the phase coexistence of ferromagnetic (FM) metallic and antiferromagnetic (AFM) insulating states via the kinetic arrest of the phase transition. The formation of mixed magnetic glass, which is susceptible to an external magnetic field ($H$), is essential to the origin of colossal magnetoresistance [5,6]. The variation of rare-earth ions in perovskite manganites also generates several types of multiferroic (MF) phases [7-10]. In medium-sized rare-earth ions, the spiral spin modulation can be stabilized, inducing ferroelectricity via antisymmetric exchange strictions with strong controllability of ferroelectric properties by external magnetic fields [7]. With a smaller radius, the crystallographic structure changes into a hexagonal structure, which represents a unique improper ferroelectricity due to structural trimerization [9]. However, the perovskite structure remains intact under high pressure and accompanies the E-type AFM phase that results in another type of the MF phase driven by symmetric exchange strictions [10].

As an extension of studies on perovskite manganites, double perovskites of $R_2CoMnO_6$ (R = La, …, Lu, and Y) have recently been explored owing to their fascinating magnetic and functional properties, such as metamagnetism [11-13], spin-glass state [14-16], exchange bias effect [17-19], magnetocaloric effect [20-22], and multiferrocity [23-27]. By replacing half the Mn ions with Co ions in perovskite manganites, a double perovskite structure is formed with $Co^{2+}$ ($S = 3/2$) and $Mn^{4+}$ ($S = 3/2$) ions, alternatively located in corner-shared octahedral environments. As the size of rare-earth ions decreases, the magnetic transition temperature ($T$) arising from the dominant $Co^{2+}$ and $Mn^{4+}$ superexchange interactions decreases from 204 K for $La_2CoMnO_6$ to 48 K for $Lu_2CoMnO_6$ [28]. In these compounds, the difficulty in attaining the impeccable alteration of $Co^{2+}$ and $Mn^{4+}$ ions naturally entails additional AFM clusters which involve another valence state of $Co^{3+}$- $Mn^{3+}$, and anti-sites of ionic disorders and/or antiphase boundaries leading to

$Co^{2+}$- $Co^{2+}$ or $Mn^{4+}$- $Mn^{4+}$ pairs. The formation of anti-sites in addition to the dominant FM order [27,29,30] of $Co^{2+}$ and $Mn^{4+}$ moments is known as the mechanism for the observed magnetic exchange bias in polycrystalline $Y_2CoMnO_6$ [19]. In $Tm_2CoMnO_6$ and $Er_2CoMnO_6$ (ECMO), the neutron diffraction studies confirm that the order of $Co^{2+}$ and $Mn^{4+}$ moments is FM and the order of $Er^{3+}/Tm^{3+}$ moments at lower temperature activates the additional ferrimagnetic (FIM) order between $Er^{3+}/Tm^{3+}$ and ferromagnetic $Co^{2+}/Mn^{4+}$ sublattices [31-33]. The FIM order exhibits an inversion of the magnetic hysteresis loop in polycrystalline ECMO [34]. In $Yb_2CoMnO_6$ and $Lu_2CoMnO_6$, the $Co^{2+}$ and $Mn^{4+}$ ions display the up-up-down-down (↑↑↓↓) spin configuration in which the ferroelectricity emerges perpendicular to the $c$-axis from the cooperative $O^{2-}$ displacements through the symmetric exchange striction [23-25]. Evidently, a scientific understanding of diverse magnetic phases and interactions is crucial for finding novel functional properties in double perovskites.

In this work, the magnetic and magnetoelectric properties of single crystals of double-perovskite ECMO were studied to reveal the characteristics corresponding to the mixed FIM and MF phases. The dominant FIM order between $Er^{3+}$ and FM $Co^{2+}/Mn^{4+}$ sublattices was identified by compensated magnetization ($M$) occurring at $T_{Comp}$ = 3.15 K. From our precise measurement of isothermal $M$ in the low $T$ regime, the inversion of the magnetic hysteresis loop was observed below $T_{Comp}$, which can be explained by the delicate balance between different magnetic moments, and qualitatively by an extended Stoner–Wohlfarth model [35-38]. The ferroelectric polarization ($P$) and dielectric constant ($\varepsilon'$) measurements demonstrated an additional inclusion of the MF phase as found in $Yb_2CoMnO_6$ and $Lu_2CoMnO_6$ [23,24]. Associated with the coexistence of FIM and MF phases, the disappearance of MF phase by an external $H$ occurs simultaneously with the metamagnetic transition, revealing exclusive characteristics of the double perovskite.

**Results and Discussion**

Figure 1(a) shows the X-ray powder diffraction pattern for the ground single crystals of double perovskite ECMO at room $T$. The crystallographic structure was refined as a monoclinic structure with the $P2_1/n$ space group. The lattice constants were found to be $a$ = 5.228 Å, $b$ = 5.594 Å, and $c$ = 7.477 Å with $\beta$ = 90.244° with good agreement factors, $\chi^2$ = 1.74, $R_p$ = 7.97, $R_{wp}$ = 6.23, and $R_{exp}$ = 4.72. The crystal structures viewed from the $a$- and $c$-axes are depicted in Figs. 1(b) and (c), respectively. $Co^{2+}$ and $Mn^{4+}$ ions are alternatingly located in corner-shared octahedral environments. The oxygen octahedral cages are strongly distorted due to the small radius of the $Er^{3+}$ ion [28].

To investigate intricate magnetic properties as anticipated in the double perovskite incorporating three different magnetic ions, the $T$-dependence of magnetic susceptibility ($\chi = M/H$) was obtained. The anisotropic $\chi$ in $H = 0.05$ kOe along ($H//c$) and perpendicular ($H \perp c$) to the $c$-axis was measured upon warming in $H$ after zero-field cooling (ZFC) and upon cooling at the same $H$ (FC), as shown in Fig. 2(a). The overall $T$-dependence of $\chi$'s for two different orientations exhibits strong magnetic anisotropy, which indicates that the spins are mainly aligned along the $c$-axis. The FM order relevant to the dominant $Co^{2+}$ and $Mn^{4+}$ superexchange interactions sets in at $T_C = 67$ K, which can be determined by the sharp anomaly in the $T$ derivative of $\chi$ in $H//c$. The $T$-dependence of heat capacity divided by the temperature ($C/T$) measured upon warming in zero $H$ also exhibits the anomaly starting from $T_C$, shown in Fig. 2(b). Upon further cooling, $C/T$ shows an abrupt increase below $T_{Er} \approx 10$ K, which corresponds to the ordering of $Er^{3+}$ moments. Below $T_{Er}$, the reversal of $\chi$ was observed in both ZFC and FC measurements (Fig. 2(a)) as a characteristic signature of a ferrimagnet [39-46].

A ferrimagnet is a substance that involves a portion of opposing magnetic moments as in antiferromagnetism, but generates a net $M$ from unequal magnetic moments in the opposite directions, thus exhibiting distinct characteristics of magnetism. The FIM interaction between $Er^{3+}$ and FM $Co^{2+}/Mn^{4+}$ sublattices generates the intriguing $T$-dependence of $\chi$ following the different sequence of measurement. In the FC measurement in $H//c$, $\chi$ increases smoothly below $T_C$ with the parallel alignment of $Co^{2+}$ and $Mn^{4+}$ moments. Upon cooling further below $T_{Er}$, the $Er^{3+}$ moments begin to align oppositely to the $Co^{2+}/Mn^{4+}$ moments, which leads to a gradual decrease in $\chi$. At lower $T$, $\chi$ intersects the zero point owing to the large moment of $Er^{3+}$ spin. On the other hand, ZFC $\chi$ shows a positive value at 2 K since the $Er^{3+}$ moments tend to orient along the $H$ direction. The decrease in the effective $Er^{3+}$ moments upon increasing $T$ results in the sign change of $\chi$. Above $T_{Er}$, the negatively magnetized $Co^{2+}/Mn^{4+}$ spins begin to flip along the applied $H$ due to thermal fluctuation, which causes another sign change of $\chi$ at 48 K. To find the compensation $T$ precisely, the thermoremanent magnetization ($M_{rem}$) [47] was measured in $H//c$ (Fig. 2(c)). At 2 K, $H = 50$ kOe was applied in $H//c$ and then turned off, and $M_{rem}$ was recorded in the absence of $H$ upon warming from 2 K. The sign reversal of $M_{rem}$ occurs at $T_{Comp}$ = 3.15 K, which manifests the FIM feature of this double perovskite compound.

The anisotropic $M$ in $H//c$ and $H\perp c$ was measured up to ±90 kOe at $T$ = 2 K, shown in Fig. 3(a). For the hysteresis loop in $H//c$, solid and dashed lines denote sweeping $H$ from +90 to −90 kOe and from −90 to +90 kOe, respectively. $M$ in $H//c$ is not saturated at +90 kOe with the magnetic moment of 17.2 $\mu_B$/f.u., but it is much larger than the moment in $H\perp c$ (8.71 $\mu_B$/f.u.), indicating the magnetic easy $c$-axis. Upon decreasing $H$ from +90 kOe, $M$ decreases smoothly until it drops precipitously below 15 kOe. At low $H$, $M$ intersects the zero point at 1 kOe and exhibits the negative remanent $M$ ($M_r$) of −1 $\mu_B$/f.u. (inset of Fig. 3(a)). Further decrease in $H$ in the negative direction induces a sharp drop in $M$ at $H_C$ = −26.5 kOe. The measurement of $M$ in $H//c$ in the opposite direction completes the inverted magnetic hysteresis loop. The inversion of the hysteresis loop in $H//c$ can be analysed by an extended Stoner–Wohlfarth model within the frame of the FIM order between $Er^{3+}$ and $Co^{2+}/Mn^{4+}$ sublattices with a different magnetic anisotropy [35-38] (see Experimental section for detail). The experimental observation of inversed magnetic hysteresis loop in ECMO suggests the considerable difference of magnetic anisotropy energies between $Er^{3+}$ and $Co^{2+}/Mn^{4+}$ moments. In our calculation, we assumed that the magnetocrystalline anisotropy energy of $Co^{2+}/Mn^{4+}$ moments is three times larger than that of $Er^{3+}$ moments. With qualitative similarity, the magnetic hysteresis loop was attained from the model, as illustrated in Fig. 3(b). Based on the result, the evolution of the spin configuration for $Er^{3+}$ and $Co^{2+}/Mn^{4+}$ ions during the sweeping of $H$ from +90 to −90 kOe in $H//c$ is schematically depicted in Fig. 3(b). The red and blue arrows indicate the effective moments of $Er^{3+}$ and $Co^{2+}/Mn^{4+}$ ions, respectively. At high $H$, the $Er^{3+}$ and $Co^{2+}/Mn^{4+}$ moments tend to be aligned in the same direction due to the dominant Zeeman energy. Upon decreasing $H$, the negative exchange coupling between $Er^{3+}$ and $Co^{2+}/Mn^{4+}$ spins accompanied by a smaller magnetocrystalline anisotropy energy and larger moment of $Er^{3+}$ ions leads to the progressive decrease in the net $Er^{3+}$ moments, followed by zero net $M$ even at a positive $H$ and negative $M_r$. Decreasing $H$ further in the negative direction induces an abrupt drop in $M$, where the $Co^{2+}/Mn^{4+}$ spins are fully reversed because the Zeeman energy of $Co^{2+}/Mn^{4+}$ sublattices overcomes the anisotropy energy. Since the change in magnitude of $M$ caused by the reversal of $Co^{2+}/Mn^{4+}$ moment at the metamagnetic transition is found to be ~9 $\mu_B$/f.u. (Fig. 3(a)), the net magnetic moment of $Co^{2+}/Mn^{4+}$ spins should be ~ 4.5 $\mu_B$/f.u., which is smaller than the summation of $Co^{2+}$ and $Mn^{4+}$ moments (6 $\mu_B$/f.u.). The smaller net magnetic moment of

$Co^{2+}/Mn^{4+}$ spins is acceptable because a small portion of $Co^{2+}/Mn^{4+}$ spins is naturally reversed during the magnetization process from +90 kOe to −26.5 kOe and antiferromagnetic exchange couplings of $Co^{2+}$- $Co^{2+}$ or $Mn^{4+}$- $Mn^{4+}$ pairs are originally included from the presence of anti-sites of ionic disorders and/or antiphase boundaries.

The close relevance of $M_r$ to $T_{Comp}$ was cautiously examined by the $T$ dependent evolution of $M_r$. The full hysteresis curves up to ±90 kOe were recorded in $H//c$ at various $T$'s. The hysteresis loops below and above $T_{Comp}$ are shown within the range of ±5 kOe in Fig. 3(c) and d, respectively. Below $T_{Comp}$, all the curves present the inverted magnetic hysteresis. Upon increasing $T$, the inverted loop becomes narrow and the magnitude of negative $M_r$ decreases linearly, resulting from the reduced net $Er^{3+}$ moments by thermal fluctuation. By crossing $T_{Comp}$, the sign of $M_r$ changes and it increases gradually with an increasing $T$.

Recently, new magnetism-driven ferroelectrics, i.e. type-II multiferroics, were found in double-perovskite $Yb_2CoMnO_6$ and $Lu_2CoMnO_6$ [23,24]. The initial polycrystalline analysis of neutron diffraction and bulk electric properties for $Lu_2CoMnO_6$ suggested that the ferroelectricity arises from the symmetric exchange striction of the ↑↑↓↓ spin chains with alternating $Co^{2+}$ and $Mn^{4+}$ charge valences [48], consistent with the Ising spin chain magnet of $Ca_3CoMnO_6$ [49]. However, studies on the single crystals of $Yb_2CoMnO_6$ and $Lu_2CoMnO_6$ revealed that the ferroelectricity emerges perpendicular to the $c$-axis below $T_C$ = 52 and 48 K, respectively. Several theoretical works provided a plausible explanation for the ferroelecticity, in which the symmetric exchange strictions along the ↑↑↓↓ spin chain with alternatingly shifted $O^{2-}$ ions generate cooperative $O^{2-}$ displacements perpendicular to the $c$-axis [50-52].

The possible formation of an additional MF phase in ECMO was examined by the $H$-dependence of $P$ obtained by integrating magnetoelectric current density ($J$), measured perpendicular to the $c$-axis ($E \perp c$) at 2 K, shown in Figs. 4(a) and (b). After poling from 100 K to 2 K in $H$ = 0 kOe and $E$ = 5.7 kV/cm, the $J$ in $H//c$ exhibits a very sharp peak with peak height of ~0.76 µA/m² at the metamagnetic transition, $H_C$ = 26.5 kOe. The corresponding $P$ value at $H$ = 0 kOe and 2 K was estimated as ~0.9 µC/m², which is only two orders of magnitude smaller than the $P$ observed in $Lu_2CoMnO_6$ and signifies the presence of a small amount of the MF phase. The tiny magnitude of $P$ at 2 K implies that the exact magnetic configuration of MF phase could hardly be identified by the neutron diffraction experiment. Upon increasing $H$, the

$P$ shows the sharp step at $H_C$ and disappears above $H_C$. The simultaneous transitions at $H_C$ for the suppression of the ferroelectricity and the reversal of $Co^{2+}/Mn^{4+}$ spins in the FIM state suggest that the small amount of the additional MF phase is strongly influenced by the dominant FIM phase. In analogy with the ferroelectricity in $Lu_2CoMnO_6$, the $P$ emerged perpendicular to the $c$ axis at $H=0$ kOe in ECMO suggests that the most plausible spin configuration of the minor MF phase would be ↑↑↓↓. The disappearance of the $P$ by applying $H$ along the $c$ axis can be explained by the change of spin configuration from the ↑↑↓↓ to ↑↑↑↑.

In Fig. 4(c), the $H$-dependence of $\varepsilon'$ in $E\perp c$ is plotted, measured in $H//c$ up to ±90 kOe at $f$ = 100 kHz and $T$ = 2 K. By sweeping $H$ between +90 to −90 kOe, the whole variation of $\varepsilon'$ is only about 1 % with strong hysteretic behaviour. The maximum values occur at $H$ = ±6 kOe, followed by the sharp transitions at $H_C$. The $\varepsilon'$ shows the rather complicated $H$ dependence in comparison with the $H$ dependence of $P$. In addition to the small portion of MF phase, the additional AFM clusters formed by anti-site disorders and antiphase boundaries in the ferromagnetic $Co^{2+}/Mn^{4+}$ sublattices would also affect the isothermal $\varepsilon'$. The complicated but tiny magnitude variation of isothermal $\varepsilon'$ may result from the intricate contributions from the small portions of MF phase and AFM clusters. For a comparison with $\varepsilon'(H)$, the $H$ derivative of isothermal $M$, $dM/dH$ at 2 K is also plotted in Fig. 4(d). The $dM/dH$ reveals the similar hysteretic variation of $\varepsilon'$. The isothermal $M$ mainly reflects the response of the FIM order between the $Er^{3+}$ and $Co^{2+}/Mn^{4+}$ moments to the external $H$, as illustrated in Fig. 3(b), but also affects strongly on the hysteretic behaviour of $\varepsilon'(H)$.

The $T$-dependence of the dielectric constant ($\varepsilon'$) and tangential loss ($\tan\delta$) is displayed in Figs. 5(a) and (b), respectively, measured perpendicular to the $c$-axis ($E\perp c$) at $f$ = 100 kHz in $H//c$ with $H$ = 0, 10, 20, and 30 kOe. At zero $H$, a small and broad peak of $\varepsilon'$ at $T_C$ = 67 K was observed in Fig. 5(a), which signifies the emergence of a small amount of MF phase. Compared to the peak height of ~15 %, normalized by the value at $T_C$ = 48 K in $Lu_2CoMnO_6$,[23] it can be estimated as only about 1 % in ECMO. Despite a small portion of the MF phase in ECMO, $T_C$ is fairly enhanced. The broad peak of $\varepsilon'$ is gradually suppressed by applying $H$ along the $c$-axis, ascribed to the change in the spin configuration from ↑↑↓↓ to ↑↑↑↑, similar to that in $Lu_2CoMnO_6$.[23] Upon decreasing $T$, $\varepsilon'$ decreases linearly until it declines faster below 20 K. The overall $T$-dependence of $\varepsilon'$ and tan$\delta$ (Fig. 5(b)) below $T_\text{C}$ appears similar to those of Lu$_2$CoMnO$_6$.

While the intrinsic coupling phenomena between magnetic and ferroelectric states in single-phase type-II multiferroics were extensively explored, detailed properties of an MF phase mixed with another magnetic phase have scarcely been revealed. The $T$ evolution of magnetoelectric effect in the mixed FIM and MF phases was examined by comparison between isothermal $P$ and $M$ at $T$'s below $T_\text{comp}$. Figures 6(a) and (b) show the $H$-dependence of $P$'s and $M$'s, respectively, in $E{\perp}c$ and $H{//}c$ at $T$ = 2, 2.25, 2.5, 2.75, and 3 K, indicating that both of $P$ and $M$ vary delicately to the change of $T$. The estimated $P$'s at 2.25 and 2.5 K were 0.79 and 0.47 μC/m$^2$, respectively. As $H$ is increased, the $P$'s are suppressed with steep steps at $H$ = 28.0 and 28.7 kOe. The initial curve of $M$ at 2, 2.25, and 2.5 K also shows the step at the same $H$ as $P$, suggesting the strong intercorrelation between FIM and MF phases. At 2.75 and 3 K, $P$ magnitudes at 0 kOe are reduced as 0.43 and 0.32 μC/m$^2$, respectively. Upon increasing $H$, the $P$'s are gradually reduced and vanish above ~37 kOe, corresponding to the overall broad feature of $M$'s. Note that $P$ above $T_\text{comp}$ could not hardly be obtained because of the almost suppressed magnitude of $J$ with a broadened feature. Figures 6(c) and (d) display isothermal $\varepsilon'$ in $E{\perp}c$ at $f$ = 100 kHz and $J$ in $H{//}c$, respectively, at $T$ = 2, 2.25, 2.5, 2.75, and 3 K. The initial curve of $\varepsilon'$ at 2 and 2.25 K indicates both a sharp peak and step-like feature at the metamagnetic transition but the $\varepsilon'$ at 2.5 K shows only a step. The sharp peak of the $J$ at 2 K shifts to higher $H$ and the peak height is reduced upon slightly increasing the $T$. The weak anomaly was observed in the $\varepsilon'$ at 2.75 and 3 K, corresponding to the disappearance of $P$. As shown in the inset of Fig. 6(d), $J$'s at 2.75 and 3 K exhibit wide and small peaks around 35 kOe.

The $T$ evolution of the magnetodielectric effect in a wide range of $T$'s in the mixed FIM and MF phases was also investigated by comparison between isothermal $\varepsilon'$ and $M$. Figure 7 displays the isothermal $\varepsilon'$ in $E{\perp}c$ at $f$ = 100 kHz and $M$ in $H{//}c$ and $H{\perp}c$, at $T$ = 5, 10, 20, 35, 50, and 65 K. At 5 K, a butterfly-like shape of $\varepsilon'$ was observed with a strong magnetic hysteresis, with the absence of the step-like metamagnetic transition (Fig. 7(a)). The broadened feature of $\varepsilon'$ is compatible with the modulation of $M$ in $H{//}c$ with the narrow magnetic hysteresis described as small values of $M_\text{r}$ = 1.22 μ$_\text{B}$/f.u. and the coercive field of $H_\text{c}$ = 2.10 kOe (Fig. 7(g)). At 10 K, the butterfly-like shape of $\varepsilon'$ is maintained (Fig. 7(b)), but the magnetic hysteresis is considerably reduced. The central part of the hysteresis loop in $H//c$ is extended as $M_r = 2.73$ $\mu_B$/f.u. and $H_c = 7.24$ kOe (Fig. 7(h)), indicative of the reduced strength of the $Er^{3+}$ spin order. In addition, the slight and elongated hysteretic behaviour of $M$ in $H \perp c$ emerges. As $T$ increases further, the magnetic hysteresis in both $\varepsilon'$ and $M$ is progressively reduced. At 65 K, just below $T_C$, the sharp cusp of $\varepsilon'$ occurs at zero $H$ with the hysteresis loop in $M$ vanishing.

In summary, we explored the magnetic and magnetoelectric properties of mixed ferrimagnetic and multiferroic phases of single-crystalline double-perovskite $Er_2CoMnO_6$. The dominant $Co^{2+}$ and $Mn^{4+}$ superexchange interactions lead to the ferromagnetic order below $T_C = 67$ K, aligned mainly along the $c$-axis. The long-range order of $Er^{3+}$ moments below $T_{Er} = 10$ K induces the ferrimagnetic order and magnetization compensation at $T_{Comp} = 3.15$ K, delicately balanced with the ferromagnetic $Co^{2+}/Mn^{4+}$ sublattice. The extended Stoner–Wohlfarth model depicts qualitatively the inverted magnetic hysteresis loop observed below $T_{Comp}$. The observation of electric polarization at low temperature is indicative of the presence of a small portion of a multiferroic phase simultaneously with the ferrimagnetic phase. The strong magnetoelectric correlation at the metamagnetic transition in the phase coexistence reveals the unique characteristic of the double perovskite compound, which offers crucial clues for exploring suitable materials for magnetoelectric functional applications.

**Methods**

Rod-shaped single crystals of ECMO with a typical size of $2 \times 2 \times 5$ mm$^3$ were grown by the conventional flux method with $Bi_2O_3$ flux in air. $Er_2O_3$, $Co_3O_4$, and $MnO_2$ powders were mixed in the stoichiometric ratio for ECMO and ground in a mortar, followed by pelletizing and calcining at 1000 °C for 12 h in a box furnace. The calcined pellet was delicately reground and sintered at 1100 °C for 24 h. The same sintering procedure after regrinding was carried out at 1200 °C for 48 h. A mixture of pre-sintered polycrystalline powder and $Bi_2O_3$ flux with a ratio of 1:12 ratio was heated to 1300 °C in a Pt crucible. It was melted at the soaking $T$ for 5 h, slowly cooled to 985 °C at a rate of 2 °C/h, and cooled to room $T$ at a rate of 250 °C/h. The crystallographic structure and absence of a second phase were checked by the Rietveld refinement [53] using the FullProf program [54] for the power X-ray diffraction data. The data were obtained with a Rigaku D/Max 2500 powder X-ray diffractometer using Cu-K$_\alpha$ radiation.

The *T* and *H* dependences of DC *M* were examined by using a VSM magnetometer in a Quantum Design PPMS (Physical Properties Measurement System). The specific heat (*C*) was measured with the standard relaxation method in PPMS. The *T* and *H* dependences of *ε′* were observed at *f* = 100 kHz using an LCR meter (E4980, Agilent). The *H* dependence of electric polarization (*P*) was obtained by the integration of magnetoelectric current measured with the *H* variation of 0.1 kOe/s after poling in a static electric field of *E* = 5.7 kV/cm.

In our extended Stoner–Wohlfarth model [35-38], the magnetic energy density can be expressed as

$$E = -M_{Er}H\cos\theta_{Er} - M_{Co/Mn}H\cos\theta_{Co/Mn} - J\cos(\theta_{Er} - \theta_{Co/Mn}) \\ + K_{Er}\sin^2(\theta_{Er} - \varphi_{Er}) + K_{Co/Mn}\sin^2(\theta_{Co/Mn} - \varphi_{Co/Mn}) \quad (1)$$

The first two terms represent the Zeeman energy density, where $M_{Er}$ and $M_{Co/Mn}$ are the effective magnetic moments of $Er^{3+}$ and $Co^{2+}/Mn^{4+}$ ions, respectively, and $\theta$ is the angle between the corresponding *M* and applied *H*. The third term denotes the exchange energy density, where the moments in the $Er^{3+}$ and $Co^{2+}/Mn^{4+}$ sublattices tend to be ordered oppositely, and thus, the exchange coupling constant (*J*) is negative. The last two terms signify the densities of magnetocrystalline anisotropy energy ($K_{Er}$ and $K_{Co/Mn}$) for $Er^{3+}$ and $Co^{2+}/Mn^{4+}$ ions, respectively, where $\varphi$ is the angle between *H* and the magnetic easy axis. The energy density can be minimized to determine the direction of net *M* at an applied *H* by solving both $\frac{\partial E}{\partial \theta_{Er}} = 0$ and $\frac{\partial E}{\partial \theta_{Co/Mn}} = 0$. The estimated net *M* can be written as follows: $M = M_{Er}\cos\theta_{Er} + M_{Co/Mn}\cos\theta_{Co/Mn}$. The calculated hysteresis loop in Figure 3(b) was obtained with $J = -4.241 \times 10^5$ J/m³, $K_{Er} = 8.482 \times 10^4$ J/m³, and $K_{Co/Mn} = 2.545 \times 10^5$ J/m³.

**Acknowledgements**

This work was supported by the NRF Grant (NRF-2016R1C1B2013709, NRF-2017K2A9A2A08000278, 2017R1A5A1014862 (SRC program: vdWMRC center), and NRF-2018R1C1B6006859).

**Author contributions**

Y.J.C. and N.L. designed the experiments. M.K.K. calculated Stoner–Wohlfarth model, and S.H.O. and J.Y.M. synthesised the single crystals. M.K.K., J.Y.M. and D.G.O. performed magnetization, heat capacity, dielectric constant, and magnetoelectric current measurements. M.K.K., Y.J.C., and N.L. analysed the data and prepared the manuscript. All the authors have read and approved the final version of the manuscript.

**Additional information**

The authors declare no competing interests.

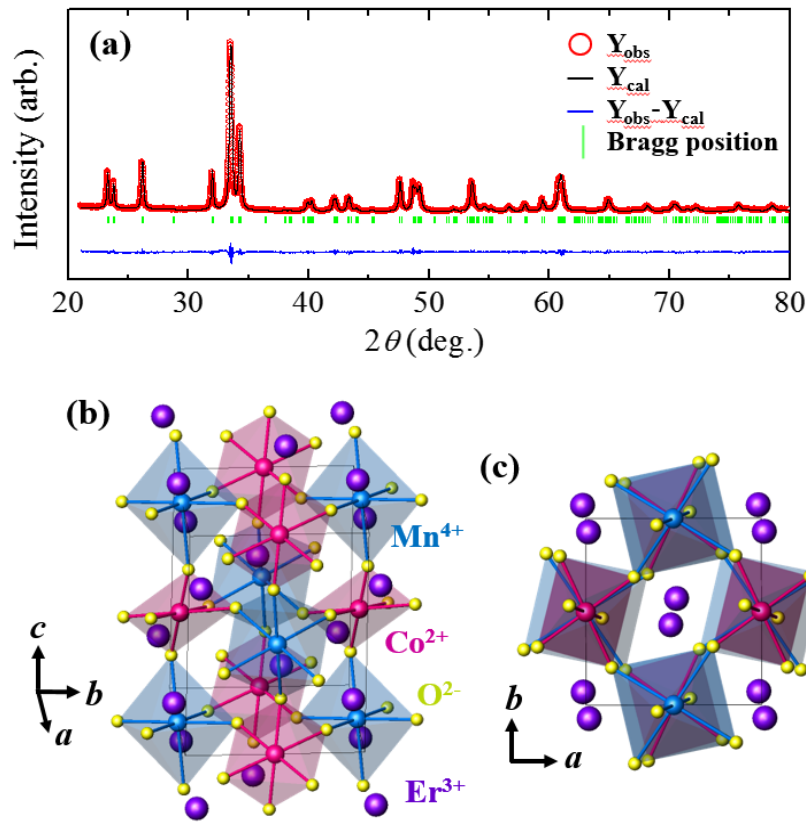

**Figure 1. Crystallographic structure of Er₂CoMnO₆.** (a) Observed (open circles) and calculated (solid line) powder X-ray diffraction patterns for ground $Er_2CoMnO_6$ (ECMO) single crystals. $Y_{obs}$, $Y_{cal}$, and $Y_{obs}-Y_{cal}$ represent the intensities of the observed patterns, calculated patterns, and their difference, respectively. The green short lines denote the Bragg positions. (b) and (c) Views of the crystal structure of double perovskite ECMO from the $a$- and $c$-axes, respectively. The purple, pink, blue, and yellow spheres represent $Er^{3+}$, $Co^{2+}$, $Mn^{4+}$, and $O^{2-}$ ions, respectively.

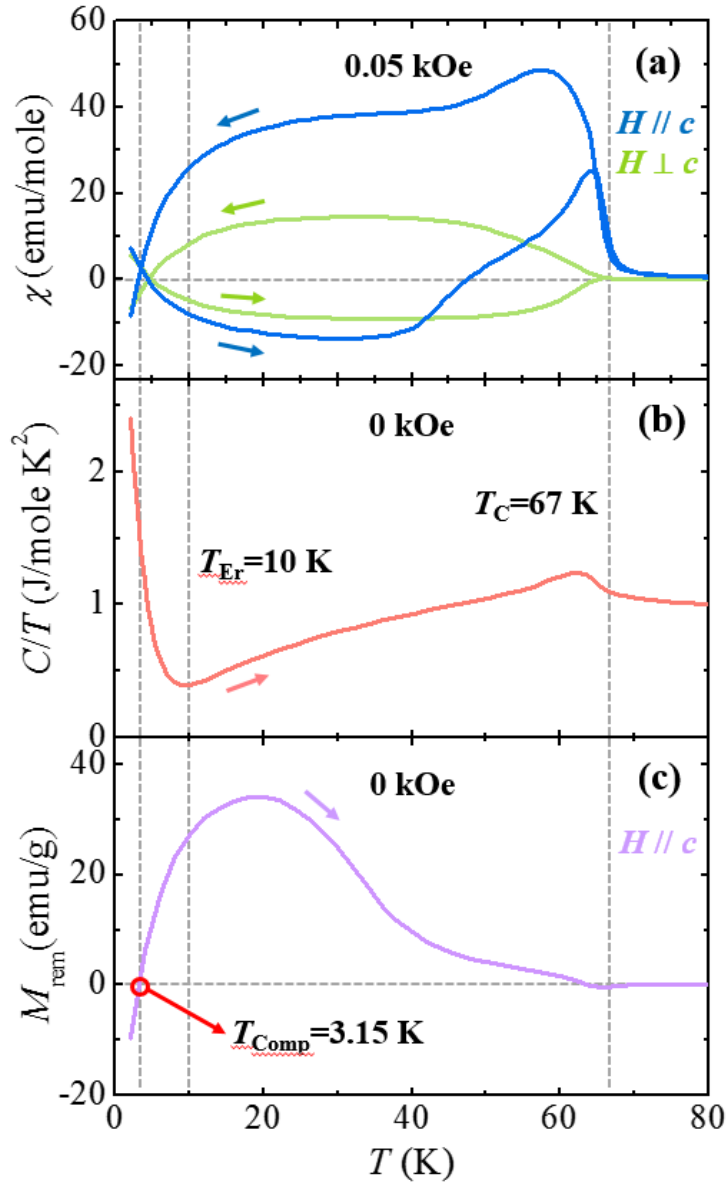

**Figure 2. Temperature-dependent magnetic properties of Er$_2$CoMnO$_6$.** (a) Temperature ($T$) dependence of the magnetic susceptibility ($\chi = M/H$, 1 emu = $4\pi \times 10^{-6}$ m$^3$) of a double-perovskite ECMO single crystal along ($H//c$) and perpendicular ($H\perp c$) to the $c$-axis, measured upon warming in $H$ = 0.05 kOe after zero-magnetic-field cooling (ZFC) and upon cooling in the same field (FC), shown up to 80 K. (b) $T$-dependence of specific heat divided by temperature ($C/T$) measured without magnetic field ($H$). (c) $T$-dependence of the thermoremanent magnetization ($M_{rem}$) of the ECMO crystal, measured in $H//c$ warming from 2 K in the absence of $H$ after cooling in 50 kOe. The vertical dashed lines indicate the ferromagnetic transition temperature ($T_C$), the Er$^{3+}$ spin ordering temperature ($T_{Er}$), and the compensation temperature ($T_{Comp}$), respectively.

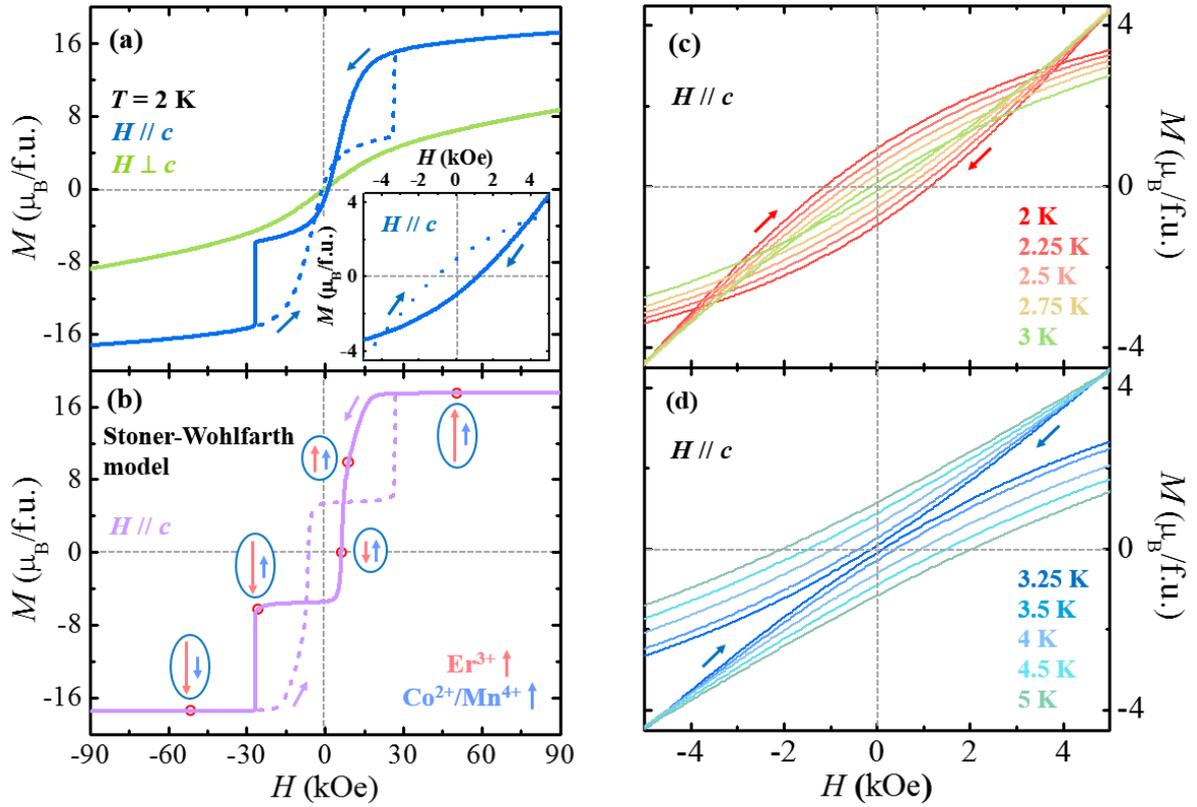

**Figure 3. Observed and calculated isothermal magnetization at 2 K and temperature evolution of inverted magnetic hysteresis.** (a) Isothermal magnetization ($M$) of the ECMO crystal in $H//c$ and $H\perp c$, measured at $T$ = 2 K up to 90 kOe after ZFC. The inset shows the magnified view in the range of $H = \pm 10$ kOe of the hysteresis loop in $H//c$. For the hysteresis loop in $H//c$, the solid and dashed curves indicate the data obtained by sweeping $H$ from +90 kOe to −90 kOe, and by sweeping $H$ from −90 kOe to +90 kOe, respectively. (b) Calculated hysteresis loop in $H//c$ by adopting the extended Stoner-Wohlfarth model. The schematic spin configurations depicted as net moments of $Er^{3+}$ (light red arrows) and $Co^{2+}/Mn^{4+}$ (light blue arrows) spins are illustrated for the curve of sweeping $H$ from +90 kOe to −90 kOe. (c) and (d) Magnified views in the range of $H = \pm 5$ kOe of isothermal $M$'s in $H//c$, measured at various $T$'s below $T_{Comp}$ ($T$ = 2, 2.25, 2.5, 2.75, and 3 K) and above $T_{Comp}$ ($T$ = 3.25, 3.5, 4, 4.5, and 5 K), respectively.

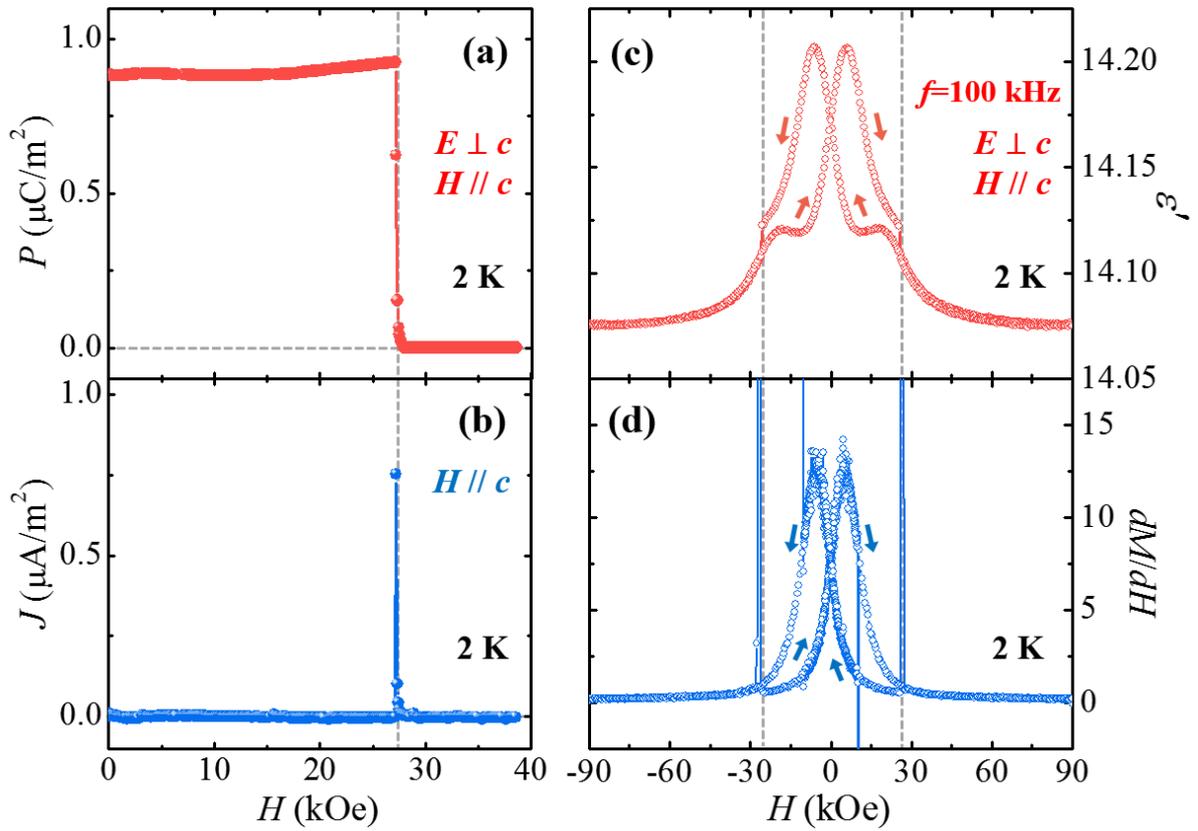

**Figure 4. Isothermal ferroelectric polarization and dielectric constant at 2 K.** (a) $H$-dependence of ferroelectric polarization ($P$) at 2 K, obtained by integrating the magnetoelectric current in b). (b) $H$-dependence of current density ($J$), measured with the $H$ variation of 0.1 kOe/s in $H//c$ after poling from 100 K to 2 K in $E$ = 5.7 kV/cm perpendicular to the $c$ axis. (c) $H$-dependence of $\varepsilon'$ in $E\perp c$, measured up to ±90 kOe in $H//c$ at 2 K. (d) $H$-derivative of isothermal $M$ ($dM/dH$) at 2 K, taken from the data in Figure 3a).

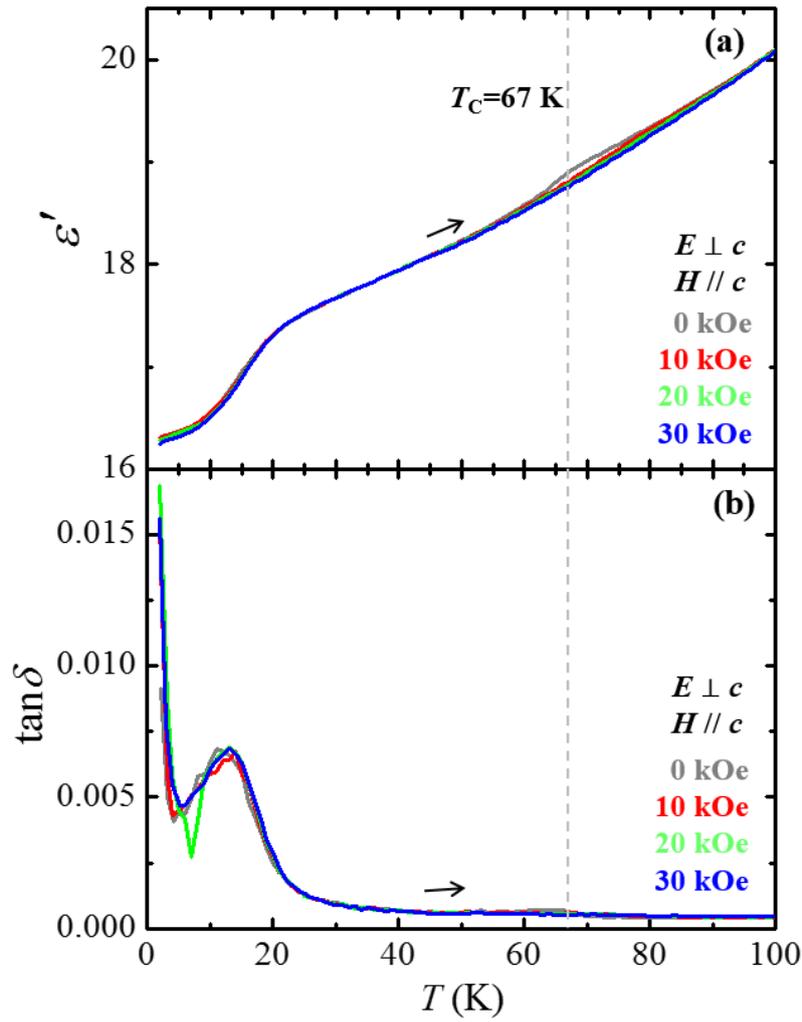

**Figure 5. Temperature dependence of the dielectric properties of Er$_2$CoMnO$_6$.** (a) and (b) $T$-dependences of dielectric constant ($\varepsilon'$) and dielectric tangential loss (tan$\delta$), respectively, measured upon warming from 2 K to 100 K in an applied AC voltage of $V$ = 1 V at $f$ = 100 kHz perpendicular to the $c$-axis ($E \perp c$), and $H$ = 0, 10, 20, and 30 kOe along the $c$-axis ($H//c$).

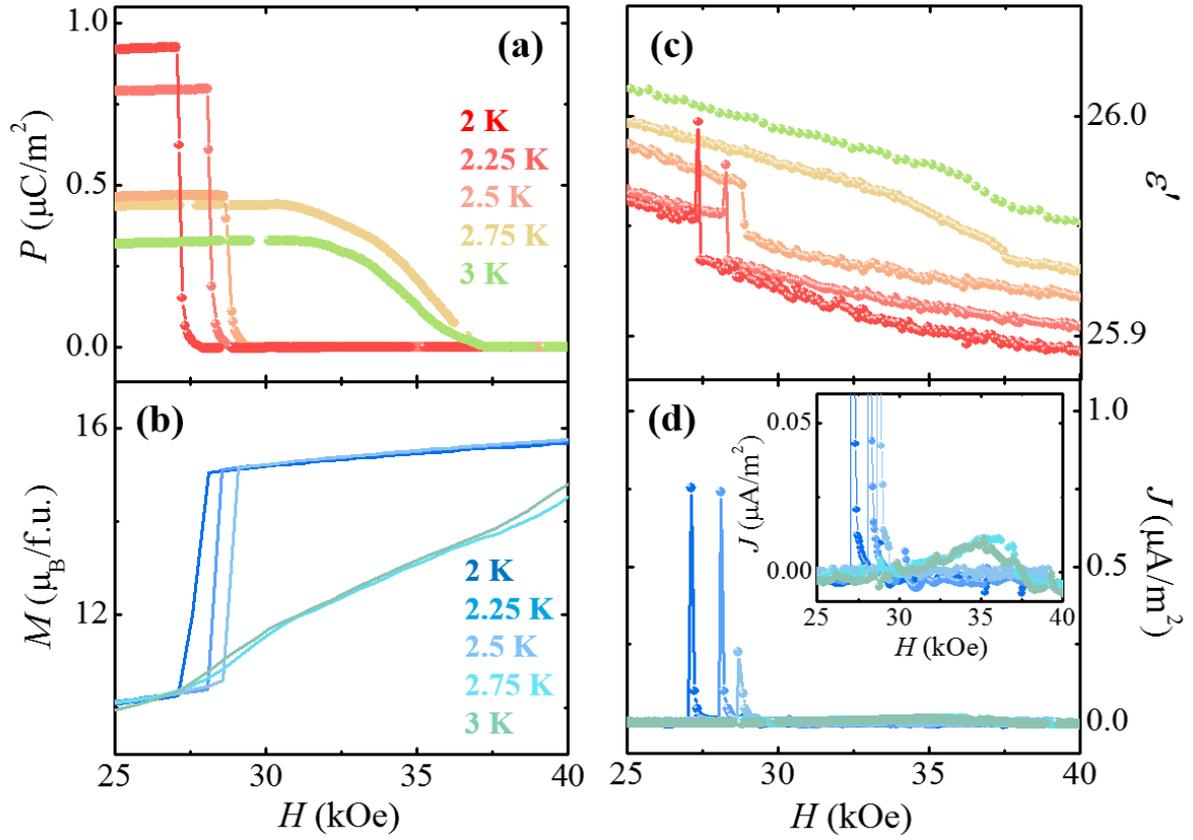

**Figure 6. Temperature evolution of ferroelectric polarization below $T_{comp}$ in comparison with that of magnetization and dielectric constant.** (a) $H$-dependence of $P$ at $T$ = 2, 2.25, 2.5, 2.75 and 3 K shown in the range of 25-40 kOe. (b) Initial curves of isothermal $M$ at $T$ = 2, 2.25, 2.5, 2.75 and 3 K. (c) $H$-dependence of $\varepsilon'$ at $T$ = 2, 2.25, 2.5, 2.75 and 3 K. (d) $H$-dependence of $J$, measured with the $H$ variation of 0.1 kOe/s in $H//c$ at $T$ = 2, 2.25, 2.5, 2.75 and 3 K after poling in $E \perp c$. The inset shows the magnified view of $J$.

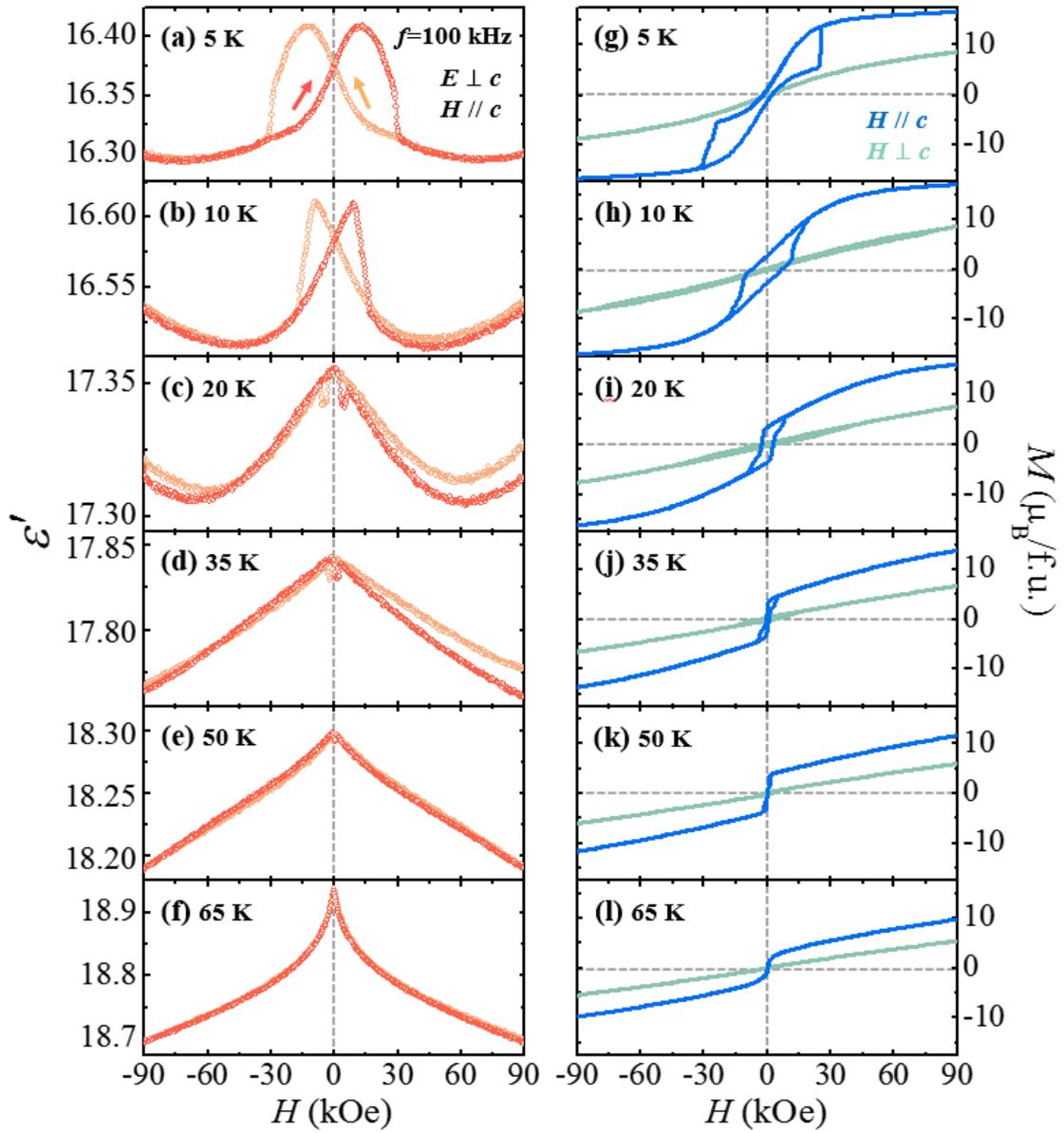

**Figure 7. Temperature evolution of the isothermal dielectric constant.** (a)-(f) Isothermal $\varepsilon'$ in $E\perp c$, measured up to ±90 kOe in $H//c$ at $T$ = 5, 10, 20, 35, 50, and 65 K, respectively. The light red and orange curves indicate the data obtained by sweeping $H$ from +90 kOe to −90 kOe, and by sweeping $H$ from −90 kOe to +90 kOe, respectively. (g)-(l) Isothermal $M$ in both $H//c$ and $H\perp c$, measured up to ±90 kOe at $T$ = 5, 10, 20, 35, 50, and 65 K, respectively.